\documentclass[endfloats, preprint,aps,pra]{revtex4}%
\usepackage{graphicx,amsmath,amsthm,amssymb}
\usepackage{amsfonts}
\usepackage{amsmath}
\usepackage{amssymb}
\usepackage{graphicx}%
\begin{document}
\title{Partially-Time-Ordered Schwinger-Keldysh Loop Expansion of Coherent Nonlinear
Optical Susceptibilities}
\author{Shaul Mukamel}
\affiliation{Department of Chemistry, University of California, Irvine, CA 92697}
\date{\today}

\begin{abstract}
A compact correlation-function expansion is developed for $n$'th order optical
susceptibilities in the frequency domain using the Keldysh-Schwinger loop.
\ By not keeping track of the relative time ordering of bra and ket
interactions at the two branches of the loop, the resulting expressions
contain only $n+1$ basic terms, compared to the 2$^{n}$ terms required for a
fully time-ordered density matrix description. \ Superoperator Green's
function expressions for $\chi^{(n)}$ derived using both expansions reflect
different types of interferences between pathways .These are demonstrated for
correlation-induced resonances in four wave mixing signals.

\end{abstract}
\maketitle


\bigskip\textbf{I.\qquad INTRODUCTION}

Time ordered expansions form the basis for the perturbative calculation of
static and dynamical properties of interacting many-body systems. \ The
nonlinear response to a sequence of $n$ short (impulsive) pulses is most
naturally calculated in real (physical) time. \ The resulting response
functions contain $2^{n}$ basic terms, stemming from the fact that each
interaction can occur either with the ket or with the bra of the system
density matrix. \ This fully time ordered expansion is routinely used for
computing ultrafast (femtosecond) optical signals in molecules, semiconductors
and other materials. \ The physical picture is recast in terms of the
density matrix in Liouville space. Many-body theory of externally driven
systems is in contrast commonly formulated using nonequilibrium Green's
functions which act in Hilbert space [1-5]. Time-ordering is then
maintained on an artificial Keldysh-Schwinger loop, [6,7] which corresponds
to both forward and backward evolution in physical time and forms the basis
for peturbative diagrammatic techniques. The loop provides a formal
bookkeeping device for various interactions. \ We only keep track of the
number of interactions with the ket and the bra but not of their relative time
ordering. \ The nonlinear response function recast using these artificial
(loop) time variables has then a considerably reduced number of terms, $n+1$.
\ Time-domain optical experiments performed using impulsive ultrashort pulses
may be described on the loop, but the required transformation from loop-to
real-time variables makes it hard to attribute physical meaning to the various terms [8].

In this paper we show that the loop time ordering is most suitable for
computing nonlinear susceptibilities in the frequency domain, where real-time
ordering is not maintained in any case. \ The frequency variables are directly
conjugated to the various delay periods along the loop. \ In Sec.II we derive
the correlation function loop expressions for the third order susceptibility.
\ Since the loop expansion is much more compact, it may be advantageous to
perform many-body calculations in the frequency domain on the loop and then
switch to the time domain by a Fourier transform. \ This way one may exploit
the full power of many body Green's function techniques. \ These expressions
are then recast in Sec.III using a diagrammatic representation in terms of
superoperators in Liouville space. \ The loop and the time-ordered expressions
are compared in Sec.IV and shown to contain a different structure of
resonances. A superficial look at the two types of expressions may suggest
that they predict different types of resonances. \ This is however misleading
since the various terms interfere. \ Consequently some apparent resonances may
cancel and others may be induced by dephasing processes. \ Simple diagrammatic
rules are provided which allow to compute the partially time-ordered
expressions. \ These subtle effects are illustrated in Sec.V by applying this
formalism to study correlation-induced resonances in four wave mixing. \ The
four point dipole correlation function is calculated for a multilevel system
whose energy levels fluctuate by coupling to a Brownian oscillator bath. \ The
model allows for an arbitrary degree of correlation between these
fluctuations. \ The expressions may not be generally factorized into products
of either real-time delays or loop-delays and the resulting complex pattern of
resonances may not be attributed to specific time delays. \ When these
fluctuations are negligible the loop expressions best reveal the resonances.
\ In the limit of fast fluctuations (homogeneous dephasing) the real time
expressions show these resonances. \ These subtle effects are demonstrated in
Sec.V where we illustrate the different role of interference in the two types
of expansion. We conclude by a discussion of these results in Sec.VI. \ 

\bigskip

\textbf{II.\qquad NONLINEAR SUSCEPTIBILITIES ON THE KELDYSH LOOP}

We consider a system interacting with an external electric optical field
$E(t)$. The coupling Hamiltonian is $H_{int}=-E(t)V,$ where V is the dipole
operator. The nonlinear polarization $P^{(n)}$ has $n+1$ terms [9,10]
\begin{equation}
P^{(n)}(t)=\sum_{m=0}^{n}\langle\psi^{(n-m)}(t)|V|\psi^{(m)}(t)\rangle.
\label{Eq:aa}%
\end{equation}
Here $\mid$ $\psi^{(m)}>$ is the perturbed wavefuntion to $m^{\prime}$th order
in the external field. We shall carry out the calculation for the third order
response, $n=3$. The generalization to n'th order is straightforward. The
field consists of three modes and expanded as%
\begin{equation}
E(t)=\sum_{j=1}^{3}E_{j}(t)\exp(-i\omega_{j}t)+c.c.
\end{equation}

\begin{figure}[p]
\includegraphics[scale=0.6]{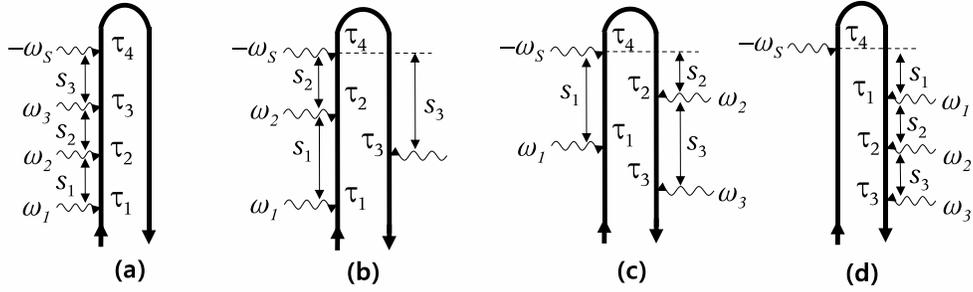}\caption{The four loop diagrams
for $\chi^{(3)}$ representing $P_{a},P_{b},P_{c}$ and $P_{d}$ in Eq.(3). The
loop expansion does not keep track of the relative time ordering of the bra
and the ket. $s_{1},s_{2}$ and $s_{3}$ are the time intervals ordered along
the loop.}%
\label{Fig:feyn1}%
\end{figure}

Eq.~(\ref{Eq:aa}) now has four terms which correspond to $m$=3,2,1,0 and are
represented by the Feynman diagrams (a), (b), (c), (d) shown in
Fig.~\ref{Fig:feyn1} respectively. The system interacts with the fields
$E_{1},E_{2}$ and $E_{3}$ at times $\tau_{1}$, $\tau_{2}$ and $\tau_{3}$
respectively, and the polarization is calculated at $\tau_{4}$\ by integrating
over the time variables $\tau_{j}.$Each diagram represents a different
ordering of $\tau_{j}$ along the loop.

Fourier transform of Eq.~(\ref{Eq:aa}) to the frequency domain gives%
\begin{equation}
P^{(3)}(\omega_{s})\equiv\int_{-\infty}^{\infty}dt\exp(i\omega_{s}%
t)P^{(3)}(t)=P_{a}+P_{b}+P_{c}+P_{d}%
\end{equation}

where%

\begin{equation}
P_{a}(\omega_{s})=\int_{-\infty}^{\infty}d\tau_{4}\int_{-\infty}^{\tau_{4}%
}d\tau_{3}\int_{-\infty}^{\tau_{3}}d\tau_{2}\int_{-\infty}^{\tau_{2}}d\tau
_{1}\text{ }E_{1}(\tau_{1})E_{2}(\tau_{2})E_{3}(\tau_{3})
\end{equation}

\[
F(\tau_{4},\tau_{3},\tau_{2},\tau_{1})\text{ }\exp(-i\omega_{1}\tau
_{1}-i\omega_{2}\tau_{2}-i\omega_{3}\tau_{3}+i\omega_{s}\tau_{4})
\]

\begin{equation}
P_{b}^{{}}(\omega_{s})=-\int_{-\infty}^{\infty}d\tau_{4}\int_{-\infty}%
^{\tau_{4}}d\tau_{2}\int_{-\infty}^{\tau_{2}}d\tau_{1}\int_{-\infty}^{\tau
_{4}}d\tau_{3}\text{ }E_{1}(\tau_{1})E_{2}(\tau_{2})E_{3}(\tau_{3})
\end{equation}

\[
F(\tau_{3},\tau_{4},\tau_{2},\tau_{1})\exp(-i\omega_{1}\tau_{1}-i\omega
_{2}\tau_{2}-i\omega_{3}\tau_{3}+i\omega_{s}\tau_{4})
\]

\begin{equation}
P_{c}^{{}}(\omega_{s})=\int_{-\infty}^{\infty}d\tau_{4}\int_{-\infty}%
^{\tau_{4}}d\tau_{1}\int_{-\infty}^{\tau_{4}}d\tau_{2}\int_{-\infty}^{\tau
_{2}}d\tau_{3}\text{ }E_{1}(\tau_{1})E_{2}(\tau_{2})E_{3}(\tau_{3})
\end{equation}

\[
F(\tau_{3},\tau_{2},\tau_{4},\tau_{1})\exp(-i\omega_{1}\tau_{1}-i\omega
_{2}\tau_{2}-i\omega_{3}\tau_{3}+i\omega_{s}\tau_{4})
\]

\begin{equation}
P_{d}^{{}}(\omega_{s})=-\int_{-\infty}^{\infty}d\tau_{4}\int_{-\infty}%
^{\tau_{4}}d\tau_{1}\int_{-\infty}^{\tau_{1}}d\tau_{2}\int_{-\infty}^{\tau
_{2}}d\tau_{3}\text{ }E_{1}(\tau_{1})E_{2}(\tau_{2})E_{3}(\tau_{3})
\end{equation}

\[
F(\tau_{3},\tau_{2},\tau_{1},\tau_{4})\exp(-i\omega_{1}\tau_{1}-i\omega
_{2}\tau_{2}-i\omega_{3}\tau_{3}+i\omega_{s}\tau_{4}).
\]

\bigskip Here we have defined the correlation function
\begin{equation}
F(\tau_{4},\tau_{3},\tau_{2},\tau_{1})\equiv(\frac{i}{\hbar})^{3}Tr\left[
V(\tau_{4})V(\tau_{3})V(\tau_{2})V(\tau_{1})\rho\right]  \equiv(\frac{i}%
{\hbar})^{3}<V(\tau_{4})V(\tau_{3})V(\tau_{2})V(\tau_{1})>, \label{Eq:ae}%
\end{equation}
where $\rho$ is the equilibrium density matrix and $V(t)$ are interaction
picture operators with respect to the free system Hamiltonian H%
\begin{equation}
V(t)=\exp(iHt)V\exp(-iHt). \label{Eq:ah}%
\end{equation}
By introducing Heavyside step functions $\theta(t)$ we can set all time
integration limits from $-\infty$ to $\infty$, and combine the four terms as%

\begin{equation}
P^{(3)}(\omega_{s})=\int_{-\infty}^{\infty}\int_{-\infty}^{\infty}%
\int_{-\infty}^{\infty}\int_{-\infty}^{\infty}d\tau_{4}d\tau_{3}d\tau_{2}%
d\tau_{1}\exp(-i\omega_{1}\tau_{1}-i\omega_{2}\tau_{2}-i\omega_{3}\tau
_{3}+i\omega_{s}\tau_{4}) \label{Eq:ab}%
\end{equation}

\[
\lbrack\theta\tau_{21})\theta(\tau_{32})\theta(\tau_{43})F(\tau_{4},\tau
_{3},\tau_{2},\tau_{1})E_{1}(\tau_{1})E_{2}(\tau_{2})E_{3}(\tau_{3})
\]

\[
-\theta(\tau_{21})\theta(\tau_{42})\theta(\tau_{43})F(\tau_{3},\tau_{4}%
,\tau_{2},\tau_{1})E_{1}(\tau_{1})E_{2}(\tau_{2})E_{3}(\tau_{3})
\]

\[
+\theta(\tau_{41})\theta(\tau_{42})\theta(\tau_{23})F(\tau_{3},\tau_{2}%
,\tau_{4},\tau_{1})E_{1}(\tau_{1})E_{2}(\tau_{2})E_{3}(\tau_{3})
\]

\[
-\theta(\tau_{41})\theta(\tau_{12})\theta(\tau_{23})F(\tau_{3},\tau_{2}%
,\tau_{1},\tau_{4})E_{1}(\tau_{1})E_{2}(\tau_{2})E_{3}(\tau_{3})
\]

The third order susceptibility $\chi^{(3)}$ is defined by [9]
\begin{equation}
P^{(3)}(\omega_{s})=\int\int\int d\omega_{1}d\omega_{2}d\omega_{3}\chi
^{(3)}(-\omega_{s};\omega_{1},\omega_{2},\omega_{3})E_{1}(\omega_{1}%
)E_{2}(\omega_{2})E_{3}(\omega_{3})\delta(\omega_{s}-\omega_{1}-\omega
_{2}-\omega_{3}), \label{Eq:ac}%
\end{equation}

where
\begin{equation}
E_{j}(\omega)=\int dt\text{ }E_{j}(t)\exp(i\omega t).
\end{equation}

By comparing Eqs.~(\ref{Eq:ab}) and (\ref{Eq:ac}) we get%
\begin{align}
\chi^{(3)}(-\omega_{s};\omega_{1},\omega_{2},\omega_{3})  &  =\frac{1}%
{(2\pi)^{2}}\sum_{p}\int_{0}^{\infty}ds_{1}\int_{0}^{\infty}ds_{2}\int
_{0}^{\infty}ds_{3}\label{Eq:af}\\
&  [F(s_{1}+s_{2}+s_{3},s_{1}+s_{2},s_{1},0)\exp(i\omega_{1}s_{1}+i(\omega
_{1}+\omega_{2})s_{2}+i(\omega_{1}+\omega_{2}+\omega_{3})s_{3})-\nonumber\\
&  F(s_{1}+s_{2}-s_{3},s_{1}+s_{2},s_{1},0)\exp(i\omega_{1}s_{1}+i(\omega
_{1}+\omega_{2})s_{2}-i(\omega_{1}-\omega_{s}+\omega_{2})s_{3})+\nonumber\\
&  F(s_{1}-s_{2}-s_{3},s_{1}-s_{2},s_{1},0)\exp(i\omega_{1}s_{1}-i(\omega
_{1}-\omega_{s})s_{2}-i(\omega_{1}-\omega_{s}+\omega_{2})s_{3})-\nonumber\\
&  F(0,s_{3},s_{3}+s_{2},s_{3}+s_{2}+s_{1})\exp(i\omega_{s}s_{1}-i(-\omega
_{s}+\omega_{1})s_{2}-i(-\omega_{s}+\omega_{1}+\omega_{2})s_{_{3}})]\nonumber
\end{align}
In Eq.(13) $s_{j}$ are the time intervals between the various interactions
along the loop. The frequency arguments of $\chi^{(3)}$ are thus naturally
connected with these variables. \ Time ordering is thus maintained on the loop
but not in real (physical) time. $\ \sum_{p}$denotes the sum over all 3!
permutations of $\omega_{1},\omega_{2},\omega_{3}$.

We shall now compare this result with the fully time-ordered expressions for
the response functions obtained by expanding the density matrix [8]. Generally
$S^{(n)}$ has $2^{n}$ terms. For $n=3$ we get%

\begin{align}
P^{(3)}(t)  &  =\int_{0}^{\infty}dt_{1}\int_{0}^{\infty}dt_{2}\int_{0}%
^{\infty}dt_{3}\\
&  S^{(3)}(t_{3},t_{2},t_{1})E_{1}(t-t_{3}-t_{2}-t_{1})E_{2}(t-t_{3}%
-t_{2})E_{3}(t-t_{3}).\nonumber
\end{align}
Unlike Eq.~(\ref{Eq:ab}), E$_{1},$ E$_{2}$ and E$_{3}$ now represent the
first, the second, and the third pulse (chronologically ordered)%
\begin{align}
S^{(3)}(t_{3},t_{2},t_{1})  &  =(\frac{i}{\hbar})^{3}\theta(t_{1})\theta
(t_{2})\theta(t_{3})\label{Eq:ad}\\
&  <[[[V(t_{1}+t_{2}+t_{3}),V(t_{2}+t_{3})],V(t_{3})],V(0)]>.\nonumber
\end{align}

Using Eq.~(\ref{Eq:ae}), Eq.~(\ref{Eq:ad}) gives%

\begin{align}
S_{{}}^{(3)}(t_{3},t_{2},t_{1})  &  =\theta(t_{1})\theta(t_{2})\theta
(t_{3})[F(t_{1},t_{1}+t_{2},t_{1}+t_{2}+t_{3},0)\label{Eq:ar}\\
&  +F(0,t_{1}+t_{2},t_{1}+t_{2}+t_{3},t_{1})\nonumber\\
&  +F(0,t_{1},t_{1}+t_{2}+t_{3},t_{1}+t_{2})\nonumber\\
&  +F(t_{1}+t_{2}+t_{3},t_{1}+t_{2},t_{1},0)]+c.c.,\nonumber
\end{align}

The third order susceptibility is finally given by%
\begin{equation}
\chi^{(3)}(-\omega_{s};\omega_{1},\omega_{2},\omega_{3})=\frac{1}{(2\pi)^{2}%
}\sum_{p}\int_{0}^{\infty}dt_{1}\int_{0}^{\infty}dt_{2}\int_{0}^{\infty}dt_{3}
\label{Eq:ag}%
\end{equation}

\[
S_{{}}^{(3)}(t_{3},t_{2},t_{1})\exp[i\omega_{1}t_{1}+i(\omega_{1}+\omega
_{2})t_{2}\text{ }+i(\omega_{1}+\omega_{2}+\omega_{3})t_{3}].
\]

The loop expression Eq.~(\ref{Eq:af}) generally has $n+1$ terms ($4$ for
$n=3$) whereas the time ordered expression Eq.~(\ref{Eq:ag}) has a much larger
number $2^{n}$ $(8$ for $n=3)$. Note that the signal frequency $\omega_{s}$
does enter explicitly in the intergrations in Eq.(13) but not in Eq.(17).
$\ t_{j}$ are intervals between successive interactions in real time and are
most convenient for impulsive techniques. \ $s_{j}$ represent intervals along
the loop and are particularly useful for frequency-domain susceptibilities.
\ This will be demonstrated next.

\textbf{III. SUPEROPERATOR EXPRESSIONS FOR SUSCEPTIBILITIES}\label{Sec: Super}

By expressing Eq.~(\ref{Eq:ae}) in terms of superoperators we can derive a
more compact Green's function expressions for the susceptibilities using a
simple diagrammatic representation. \ Below we briefly survey the basic
elements of the Liouville space superoperator formalism [8,11-13]. With each
ordinary (Hilbert space) operator, $Q$, we associate two superoperators,
denoted as $Q_{L}$ (left) and $Q_{R}$ (right) defined through their left or
right action on some Hilbert space operator $X$,%
\begin{equation}
Q_{L}X\equiv QX,~~~~~~~~Q_{R}X\equiv XQ.
\end{equation}
We further define the linear combinations of these superoperators $Q^{+}%
$$\equiv$$\left(  Q_{L}+Q_{R}\right)  /2$ and $Q^{-}$ $\equiv$$Q_{L}$ $-Q_{R}%
$. Thus a $+$($-$) operation in Liouville space corresponds to an
anticommutation (commutation) operation in Hilbert space, $Q^{+}%
X\equiv(QX+XQ)/2$ and $Q^{-}X\equiv$$QX-XQ$.

The interaction picture for superoperators is defined by%
\begin{equation}
U_{\alpha}(t)=\exp(iLt)U_{\alpha}\exp(-iLt)\text{
\ \ \ \ \ \ \ \ \ \ \ \ \ \ \ \ \ \ }\alpha=L,R\text{\ \ \ } \label{Eq:ai}%
\end{equation}
where $LA\equiv\lbrack H,A]$ is the Liouville operator.

When substituting Eq.~(\ref{Eq:ah}) in Eq.~(\ref{Eq:ae}) we get%
\begin{equation}
F(\tau_{4},\tau_{3},\tau_{2},\tau_{1})=(\frac{i}{\hbar})^{3}Tr\left[
\exp(-iH\tau_{14})V\exp(-iH\tau_{43})V\exp(-iH\tau_{32})V\exp(-iH\tau
_{21})V\rho\right]  \text{.} \label{Eq:al}%
\end{equation}
Note that all interactions in Eq.~(\ref{Eq:ae}) are from the left i.e. they
act on the ket of the density matrix. We can thus recast it using "left"
superoperators as follows \
\begin{equation}
F(\tau_{4},\tau_{3},\tau_{2},\tau_{1})=(\frac{i}{\hbar})^{3}Tr\left[
V_{L}(\tau_{4})V_{L}(\tau_{3})V_{L}(\tau_{2})V_{L}(\tau_{1})\rho\right]  .
\label{Eq:aj}%
\end{equation}
Combining Eqs.~(\ref{Eq:ai}) and (\ref{Eq:al}) gives%
\begin{align*}
F(\tau_{4},\tau_{3},\tau_{2},\tau_{1})  &  =\\
&  (\frac{i}{\hbar})^{3}Tr\left[  \exp(iL\tau_{4})V_{L}\exp(-iL\tau_{43}%
)V_{L}\exp(-iL\tau_{32})V_{L}\exp(-iL\tau_{21})V_{L}\exp(-iL\tau_{1}%
)\rho\right]
\end{align*}

When exp$(-iL\tau_{1})$ acts on the equilibrium density matrix $\rho$ it does
not affect it and gives $\rho.$Similarly $\exp(iL\tau_{4}),$ when acts to the
left will give 1 under the trace. These two propagators can thus be dropped
and we finally get%
\begin{equation}
F(\tau_{4},\tau_{3},\tau_{2},\tau_{1})=(\frac{i}{\hbar})^{3}Tr\left[
V_{L}\exp(-iL\tau_{43})V_{L}\exp(-iL\tau_{32})V_{L}\exp(-iL\tau_{21})V_{L}%
\rho\right]  \label{Eq:ak}%
\end{equation}

Using Eq.~(\ref{Eq:ak}), the correlation functions in Eq.~(\ref{Eq:af}) now
become%
\begin{align}
F(s_{1}+s_{2}+s_{3},s_{1}+s_{2},s_{1},0)  &  =<V_{L}\mathcal{G}(s_{3}%
)V_{L}\mathcal{G}(s_{2})V_{L}\mathcal{G}(s_{1})V_{L}>\label{Eq:am}\\
F(s_{1}+s_{2}-s_{3},s_{1}+s_{2},s_{1},0)  &  =<V_{L}\mathcal{G}^{\dag}%
(s_{3})V_{L}\mathcal{G}(s_{2})V_{L}\mathcal{G}(s_{1})V_{L}>\nonumber\\
F(s_{1}-s_{2}-s_{3},s_{1}-s_{2},s_{1},0)  &  =<V_{L}\mathcal{G}^{\dag}%
(s_{3})V_{L}\mathcal{G}^{\dag}(s_{2})V_{L}\mathcal{G}(s_{1})V_{L}>\nonumber\\
F(0,s_{3},s_{3}+s_{2},s_{3}+s_{2}+s_{1})  &  =<V_{L}\mathcal{G}^{\dag}%
(s_{3})V_{L}\mathcal{G}^{\dag}(s_{2})V_{L}\mathcal{G}^{\dag}(s_{1}%
)V_{L}>.\nonumber
\end{align}

Here we have made use of the fact that all s variables are positive and
represent "forward" propagation along the loop.%

\begin{align}
\mathcal{G}(s)  &  =(-\frac{i}{\hbar})\theta(s)\exp(-iLs-\eta s)\\
\mathcal{G}^{\dag}(s)  &  =(\frac{i}{\hbar})\theta(s)\exp(iLs-\eta s)\nonumber
\end{align}

We reiterate that ordering on the loop does not represent ordering in real
time. \ Using superoperators we were able to recast F in terms of three
propagators (in Hilbert space F, Eq.~(\ref{Eq:al}) has 4 propagators). By
substituting Eq.~(\ref{Eq:am}) in Eq.~(\ref{Eq:af}) we obtain%
\begin{align}
\chi^{(3)}(-\omega_{s};\omega_{1},\omega_{2},\omega_{3})  &  =-\frac{1}%
{(2\pi)^{2}}\sum_{p}\label{Eq:an}\\
&  <V_{L}\mathcal{G}(\omega_{1}+\omega_{2}+\omega_{3})V_{L}\mathcal{G}%
(\omega_{1}+\omega_{2})V_{L}\mathcal{G}(\omega_{1})V_{L}>\nonumber\\
-  &  <V_{L}\mathcal{G}^{\dag}(-\omega_{s}+\omega_{1}+\omega_{2}%
)V_{L}\mathcal{G}(\omega_{1}+\omega_{2})V_{L}\mathcal{G}(\omega_{1}%
)V_{L}>\nonumber\\
+  &  <V_{L}\mathcal{G}^{\dag}(-\omega_{s}+\omega_{1}+\omega_{2}%
)V_{L}\mathcal{G}^{\dag}(-\omega_{s}+\omega_{1})V_{L}\mathcal{G}(\omega
_{1})V_{L}>\nonumber\\
-  &  <V_{L}\mathcal{G}^{\dag}(-\omega_{s}+\omega_{1}+\omega_{2}%
)V_{L}\mathcal{G}^{\dag}(-\omega_{s}+\omega_{1})V_{L}\mathcal{G}^{\dag
}(-\omega_{s})V_{L}>.\nonumber
\end{align}

Here%

\begin{equation}
\mathcal{G}(\omega)=\frac{1}{\omega-L+i\eta},\text{ }%
\end{equation}

is the retarded Green's function, and%

\begin{equation}
\mathcal{G}^{^{\dag}}(\omega)=\frac{1}{\omega-L-i\eta}%
\end{equation}

is the advanced Green's function, where $\eta$ is a positive infinitesimal.

Eq.(25) may be represented by the unfolded loop diagrams shown in Fig. 2.
\ These diagrams may be constructed using the following rules:

(i) \ Each $V_{L}$ is represented by an arrow acting on the ket from the left.

(ii) \ Each $V_{L}$ is associated with one of the frequencies $\pm\omega
_{1},\pm\omega_{2},\pm\omega_{3}\pm\omega_{s}$. \ Positive frequency $+\omega$
(negative frequency $-\omega$) is represented by an arrow pointing to the
right (left).

(iii) \ There are $(\emph{n}+1)$ choices for the position of $\omega_{s}$
along the loop see Eq.~(\ref{Eq:aa}). \ Each gives one diagram.

(iv) Each interval "before" ("after") $\omega_{s}$ gives a Green's function
$\mathcal{G(\omega)}$, $(\mathcal{G}^{\dag}(\omega))$.

(v) \ The frequency argument of each Green's function is the sum of all
"earlier" frequencies along the loop (frequency is cumulative).

(vi) \ All $\omega_{j}$ other than $\omega_{s}$ can be interchanged, giving
$n!$ permutations of $\omega_{1}...\omega_{n}$ . Altogether $\chi^{(3)}$
finally has $(n+1)!$ terms.

Finally, for comparison, using the fully time-ordered expansion
Eq.~(\ref{Eq:ad}) we have%
\begin{align}
\chi^{(3)}(-\omega_{s};\omega_{1},\omega_{2},\omega_{3})  &  =-\frac{1}%
{(2\pi)^{2}}\sum_{p}\label{Eq:ao}\\
&  [\left\langle V_{L}\mathcal{G}(\omega_{1}+\omega_{2}+\omega_{3}%
)V_{R}\mathcal{G}(\omega_{1}+\omega_{2})V_{R}\mathcal{G}(\omega_{1}%
)V_{L}\right\rangle \nonumber\\
&  +\left\langle [V_{L}\mathcal{G}(\omega_{1}+\omega_{2}+\omega_{3}%
)V_{R}\mathcal{G}(\omega_{1}+\omega_{2})V_{L}\mathcal{G}(\omega_{1}%
)V_{R}\right\rangle \nonumber\\
&  +\left\langle V_{L}\mathcal{G}(\omega_{1}+\omega_{2}+\omega_{3}%
)V_{L}\mathcal{G}(\omega_{1}+\omega_{2})V_{R}\mathcal{G}(\omega_{1}%
)V_{R}\right\rangle \nonumber\\
&  +\left\langle V_{L}\mathcal{G}(\omega_{1}+\omega_{2}+\omega_{3}%
)V_{L}\mathcal{G}(\omega_{1}+\omega_{2})V_{L}\mathcal{G}(\omega_{1}%
)V_{L}\right\rangle \nonumber\\
&  -\left\langle V_{L}\mathcal{G}(\omega_{1}+\omega_{2}+\omega_{3}%
)V_{L}\mathcal{G}(\omega_{1}+\omega_{2})V_{R}\mathcal{G}(\omega_{1}%
)V_{L}\right\rangle \nonumber\\
&  -\left\langle V_{L}\mathcal{G}(\omega_{1}+\omega_{2}+\omega_{3}%
)V_{L}\mathcal{G}(\omega_{1}+\omega_{2})V_{L}\mathcal{G}(\omega_{1}%
)V_{R}\right\rangle \nonumber\\
&  -\left\langle V_{L}\mathcal{G}(\omega_{1}+\omega_{2}+\omega_{3}%
)V_{R}\mathcal{G}(\omega_{1}+\omega_{2})V_{L}\mathcal{G}(\omega_{1}%
)V_{L}\right\rangle \nonumber\\
&  -\left\langle V_{L}\mathcal{G}(\omega_{1}+\omega_{2}+\omega_{3}%
)V_{R}\mathcal{G}(\omega_{1}+\omega_{2})V_{R}\mathcal{G}(\omega_{1}%
)V_{R}\right\rangle ].\nonumber
\end{align}

Eq.(28) may be represented by the double sided Feyman diagrams [8] shown in
Fig.3. \ Note that this expression only contains retarded Green's functions
representing forward time evolution, whereas Eq.(26) contains both retarded
and advanced Green's functions. \ A more detailed comparison will be given in
the next section.

\bigskip

\textbf{IV. \ \ RESONANCE STRUCTURE AND INTERFERENCE IN THE FULLY AND
PARTIALLY TIME ORDERED EXPANSIONS}\label{Sec:Corr copy(1)}

Consider a multilevel system a, b, c... interacting with a bath whose
Hamiltonian depends on the state of the system. \ The total eigenstates in the
joint system + bath space are denoted $\mid a\alpha>,\mid$ $b\beta>,\mid
c\gamma>$ etc. Note that the manifolds $\{\alpha\},\{\beta\},\{\gamma\}$
diagonalize different bath Hamiltonians and they are not orthogonal to each
other. \ For this model the total Green's function is given by%
\begin{equation}
\mathcal{G(\omega)=}\sum_{a\alpha,b\beta}\frac{\mid a\alpha,b\beta
>><<a\alpha,b\beta\mid}{\omega-\omega_{ab}-\omega_{\alpha\beta}+i\eta}%
\end{equation}
where $\omega_{ab}\equiv\omega_{a}-\omega_{b}$ is the transition frequency
between states $|a\rangle$ and $|b\rangle$ and $\omega_{\alpha\beta}%
\equiv\varepsilon_{\alpha}-\varepsilon_{\beta}.$ \ We next define the reduced
bath Green's function by a partial trace over the system (denoted by a
subscript s)%
\begin{equation}
\mathcal{G}_{ab}^{{}}(\omega)\equiv<<ab\mid\mathcal{G(\omega)\mid
}ab\mathcal{>>}_{s}%
\end{equation}

$\mathcal{G}_{ab}^{{}}(\omega)$ is thus a superoperator in bath space%
\begin{equation}
\mathcal{G}_{ab}^{{}}(\omega)\sum_{\alpha,\beta}\frac{\mid\alpha
\beta>><<\alpha\beta\mid}{\omega-\omega_{ab}-\omega_{\alpha\beta}+i\eta}%
\end{equation}

Expanding Eq.~(\ref{Eq:an}) in eigenstates gives%
\begin{align}
&  \chi^{(3)}(-\omega_{s};\omega_{1},\omega_{2},\omega_{3})=-\frac{1}%
{(2\pi)^{2}}\sum_{a,b,c,d}P(a)V_{ad}V_{dc}V_{cb}V_{ba}\label{Eq:ap}\\
\lbrack &  <\mathcal{G}_{da}^{{}}(\omega_{1}+\omega_{2}+\omega_{3}%
)\mathcal{G}_{ca}^{{}}(\omega_{1}+\omega_{2})\mathcal{G}_{ba}^{{}}(\omega
_{1})>\nonumber\\
-  &  <\mathcal{G}_{da}^{\dag}(-\omega_{s}+\omega_{1}+\omega_{2}%
)\mathcal{G}_{ca}(\omega_{1}+\omega_{2})\mathcal{G}_{ba}(\omega_{1}%
)>\nonumber\\
+  &  <\mathcal{G}_{da}^{\dag}(-\omega_{s}+\omega_{1}+\omega_{2}%
)\mathcal{G}_{ca}^{\dag}(-\omega_{s}+\omega_{1})\mathcal{G}_{ba}(\omega
_{1})>\nonumber\\
-  &  <\mathcal{G}_{da}^{\dag}(-\omega_{s}+\omega_{1}+\omega_{2}%
)\mathcal{G}_{ca}^{\dag}(-\omega_{s}+\omega_{1})\mathcal{G}_{ba}^{\dag
}(-\omega_{_{1}})>]\nonumber
\end{align}
Here $P(a)$ is the equilibrium population of state $|a\rangle$. For
comparison, by expanding the time-ordered expression Eq.~(\ref{Eq:ao}) in
eigenstates we get%
\begin{align}
\chi^{(3)}(-\omega_{s};\omega_{1},\omega_{2},\omega_{3})  &  =-\frac{1}%
{(2\pi)^{2}}\sum_{p}\sum_{a,b,c,d}P(a)V_{ad}V_{dc}V_{cb}V_{ba} \label{Eq:long}%
\\
&  [\left\langle \mathcal{G}_{dc}(\omega_{1}+\omega_{2}+\omega_{3}%
)\mathcal{G}_{db}(\omega_{1}+\omega_{2})\mathcal{G}_{da}(\omega_{1}%
)\right\rangle \nonumber\\
&  +\left\langle \mathcal{G}_{dc}(\omega_{1}+\omega_{2}+\omega_{3}%
)\mathcal{G}_{db}(\omega_{1}+\omega_{2})\mathcal{G}_{ab}(\omega_{1}%
)\right\rangle \nonumber\\
&  +\left\langle \mathcal{G}_{dc}(\omega_{1}+\omega_{2}+\omega_{3}%
)\mathcal{G}_{ac}(\omega_{1}+\omega_{2})\mathcal{G}_{ab}(\omega_{1}%
)\right\rangle \nonumber\\
&  +\left\langle \mathcal{G}_{ba}(\omega_{1}+\omega_{2}+\omega_{3}%
)\mathcal{G}_{ca}(\omega_{1}+\omega_{2})\mathcal{G}_{da}(\omega_{1}%
)\right\rangle \nonumber\\
&  -\left\langle \mathcal{G}_{cb}(\omega_{1}+\omega_{2}+\omega_{3}%
)\mathcal{G}_{db}(\omega_{1}+\omega_{2})\mathcal{G}_{da}(\omega_{1}%
)\right\rangle \nonumber\\
&  -\left\langle \mathcal{G}_{dc}(\omega_{1}+\omega_{2}+\omega_{3}%
)\mathcal{G}_{db}(\omega_{1}+\omega_{2})\mathcal{G}_{da}(\omega_{1}%
)\right\rangle \nonumber\\
&  -\left\langle \mathcal{G}_{cb}(\omega_{1}+\omega_{2}+\omega_{3}%
)\mathcal{G}_{ca}(\omega_{1}+\omega_{2})\mathcal{G}_{da}(\omega_{1}%
)\right\rangle \nonumber\\
&  -\left\langle \mathcal{G}_{ad}(\omega_{1}+\omega_{2}+\omega_{3}%
)\mathcal{G}_{ac}(\omega_{1}+\omega_{2})\mathcal{G}_{ab}(\omega_{1}%
)\right\rangle ]\nonumber
\end{align}

It is interesting to note that Eq.(32) only suggests resonances with
transitions involving the initial state $a$, $\omega_{\nu a}$, whereas Eq.(33)
shows explicitly resonances between any pair of levels $\omega_{\nu\nu
^{\prime}}.$ \ Both expressions are however formally exact and these apparent
differences disappear by interference effects between various terms that can
cancel some apparent resonances or induce new ones. \ When the dynamics of
fluctuations is such that the loop time variables $s_{1},s_{2},s_{3}$ are
independent, the products of Green's functions in Eq.~(\ref{Eq:am}) can be
factorized and \ Eq.~(\ref{Eq:ap}) then provides a natural representation for
the observed resonances. \ Similarly, when the physical time delays between
pulses $t_{1,}t_{2}$ and $t_{3}$ are independent, products of the
corresponding Green's functions can be factorized, and Eq.~(\ref{Eq:long})
should show the proper resonances. Most generally, neither factorization holds
and averages of products of Green's functions must be carefully carried out.
This will be illustrated in the next section using a model of multilevel
system coupled to a Brownian oscillator bath.

\bigskip

\textbf{V. \ \ CORRELATION-INDUCED RESONANCES IN FOUR WAVE MIXING}

We consider a multilevel system coupled to a harmonic bath and described by
the Hamiltonian,
\begin{equation}
\hat{H}=\hat{H}_{S}+\hat{H}_{B}+\hat{H}_{SB}%
\end{equation}
where the three terms represent respectively the system, the bath, and their
interaction.
\begin{equation}
\hat{H}_{S}=\sum_{\nu}\varepsilon_{\nu}|\nu\rangle\langle\nu|,
\end{equation}
where $\varepsilon_{\nu}$ is the energy of eigenstate $\nu$.

The system is linearly coupled to the bath through
\begin{equation}
\hat{H}_{SB}=\sum_{\nu}\hat{Q}_{\nu^{{}}}|\nu\rangle\langle\nu^{{}}|,
\end{equation}
where $\hat{Q}_{\nu}$ is a collective bath coordinate which modulates the
energy of state $v$.

The response function for this model of diagonal fluctuations can be
calculated exactly using the second order cumulant expansion. Expanding the
four point correlation function in the system eigenstates we get [14]
\begin{equation}
F(\tau_{4},\tau_{3},\tau_{2},\tau_{1})=(\frac{i}{\hbar})^{3}\sum_{cba}%
V_{ad}V_{dc}V_{cb}V_{ba}\exp\left[  -i(\varepsilon_{d}\tau_{43}+\varepsilon
_{c}\tau_{32}+\varepsilon_{b}\tau_{21})+f_{dcba}(\tau_{4},\tau_{3},\tau
_{2},\tau_{1})\right]  \label{Eq:as}%
\end{equation}
where%

\begin{align}
f_{dcba}^{{}}(\tau_{4},\tau_{3},\tau_{2},\tau_{1})  &  =-g_{dd}(\tau
_{43})-g_{cc}(\tau_{32})-g_{bb}(\tau_{21})-g_{dc}(\tau_{42})\label{Eq:av}\\
&  +g_{dc}(\tau_{43})+g_{dc}(\tau_{32})-g_{db}(\tau_{41})+g_{db}(\tau
_{42})\nonumber\\
&  +g_{db}(\tau_{31})-g_{db}(\tau_{32})-g_{cb}(\tau_{31})+g_{cb}(\tau
_{32})+g_{cb}(\tau_{21}).\nonumber
\end{align}
Here
\begin{equation}
g_{\nu\nu^{\prime}}(t)=\int_{0}^{t}d\tau_{2}\int_{0}^{\tau_{2}}d\tau_{1}%
C_{\nu\nu^{\prime}}(\tau_{2}-\tau_{1}),
\end{equation}
is the line broadening function, and
\begin{equation}
C_{\nu\nu^{\prime}}(\tau_{2}-\tau_{1})\equiv\langle Q_{\nu}(\tau_{2}%
)Q_{\nu^{\prime}}(\tau_{1})\rangle
\end{equation}
is the cross correlation function of frequency fluctuations of levels $\nu$
and $\nu^{\prime}$. We note the symmetry $g_{\nu\nu^{\prime}}(t)=g_{\nu
^{\prime}\nu}^{\ast}(-t).$ Upon the substitution of these results in
Eqs.~(\ref{Eq:af}) or (\ref{Eq:ar}) and (\ref{Eq:ag}) we can calculate
$\chi^{(3)}.$ Using the Brownian oscillator model for the correlations
function we have [8]%
\begin{equation}
g_{\nu\nu^{\prime}}(t)=(\frac{2\lambda_{\nu\nu^{\prime}}kT}{\Lambda_{\nu
\nu^{\prime}}^{2}}-i\frac{\lambda_{\nu\nu^{\prime}}}{\Lambda_{\nu\nu^{\prime}%
}})(\exp(-\Lambda_{\nu\nu^{\prime}}\left\vert t\right\vert -1+\Lambda_{\nu
\nu^{\prime}}\left\vert t\right\vert ).
\end{equation}

Here $\lambda_{_{\nu\nu^{\prime}}}$ represents the coupling strength\ (the
variances of frequency fluctuations are $2\lambda_{\nu\nu^{\prime}}kT$) and
$\Lambda_{\nu\nu^{\prime}}$ is the inverse timescale of bath fluctuations.
\ For fast fluctuations $\Lambda^{2}>>2\lambda kT$ we have%

\begin{equation}
g_{_{\nu\nu^{\prime}}}(t)=\Gamma_{_{\nu\nu^{\prime}}}\left\vert t\right\vert
\label{Eq:at}%
\end{equation}

with $\Gamma_{_{\nu\nu^{\prime}}}$ $\equiv$ $2\lambda_{_{\nu\nu^{\prime}}%
}kT/\Lambda_{_{\nu\nu^{\prime}}}.$ \ In the opposite limit of slow
fluctuations $\Lambda^{2}<<2\lambda kT$ we get%
\begin{equation}
g_{\nu\nu^{\prime}}(t)=\lambda_{_{\nu\nu^{\prime}}}\text{ }kTt^{2}%
-i\lambda_{_{\nu\nu^{\prime}}}\left\vert t\right\vert
\end{equation}
Eq.(37) implies that Eq.~(\ref{Eq:am}) may not be generally factorized into
three factors that depend on $s_{1},s_{2}$ and $s_{3}$. Similarly
Eq.(\ref{Eq:ar}) may not be factorized into factors that depend on t$_{1}$,
t$_{2}$, and t$_{3}$; a three fold integration will be required to calculate
$\chi^{(3)}$ in either representation. \ 

As an example for a dramatic interference effect related to these
factorizations, let us consider the level system with a ground state a and two
closely lying excited states b and d. \ b and d can represent, for example,
two vibrational states belonging to the same electronically excited state or
two Zeeman levels. The transition dipole only connects a with b and a with d.
\ We look for two-photon resonances of the form $(\omega_{1}-\omega_{2}%
-\omega_{bd})$ in $\chi^{(3)}(-\omega_{s};\omega_{3},-\omega_{2},\omega_{1})$.
\ These are kind of Raman resonances but for excited state frequencies
$\omega_{bd}$. \ For simplicity we tune $\omega_{3}$ to be off resonant and
assume the fast fluctuation limit Eq.~(\ref{Eq:at}). In this case the Green's
functions assume the form
\[
\mathcal{G}_{\nu\nu^{\prime}}(\omega)=\frac{1}{\omega-\omega_{\nu\nu^{\prime}%
}+i\Gamma_{\nu\nu^{\prime}}}%
\]

The two diagrams responsible for such resonances in the time ordered expansion
are shown in Fig.~4. \ They give the following contribution to $\chi^{(3)}$
\begin{equation}
\mid V_{ab}\mid^{2}\mid V_{ad}\mid^{2}\frac{1}{-\omega_{s}-\omega_{ba}%
+i\Gamma_{bb}}\text{ }\frac{1}{\omega_{1}-\omega_{2}-\omega_{bd}+i\Gamma_{bd}%
}\text{ }\left(  \frac{1}{\omega_{1}-\omega_{ba}+i\Gamma_{bb}}+\frac
{1}{-\omega_{2}-\omega_{ad}+i\Gamma_{dd}}\right)  \label{Eq:au}%
\end{equation}

Here%
\begin{align}
\Gamma_{bd}  &  =\left(  \Gamma_{bb}+\Gamma_{dd}\right)  \left(  1-\eta\right)
\label{Eq:ay}\\
\lambda_{bd}  &  =\lambda_{dd}\lambda_{bb}(1-\eta) \label{Eq:bb}%
\end{align}
$\eta$ is the correlation coefficient for fluctuations of $\omega_{ba}$ and
$\omega_{da}$. \ $\eta=-1,0,1$ represent fully anticorrelated, uncorrelated
and fully correlated fluctuations. \ The two terms in the brackets can be
combined to give%
\begin{equation}
\frac{\omega_{1}-\omega_{2}-\omega_{bd}+i(\Gamma_{bb}+\Gamma_{dd})(1-\eta
+\eta)}{\left(  \omega_{1}-\omega_{ba}+i\Gamma_{bb}\right)  \left(
-\omega_{2}-\omega_{ad}+i\Gamma_{dd}\right)  }%
\end{equation}

Substituting this in Eq.(\ref{Eq:au}) results in%
\[
\mid V_{ab}\mid^{2}\mid V_{ad}\mid^{2}\frac{1}{-\omega_{s}-\omega_{ba}%
+i\Gamma_{bb}}\frac{1}{\omega_{1}-\omega_{ba}+i\Gamma_{bb}}\frac{1}%
{-\omega_{2}-\omega_{ad}+i\Gamma_{dd}}%
\]

\begin{equation}
\left[  1+\eta\frac{1}{\omega_{1}-\omega_{2}-\omega_{bd}+i(\Gamma_{bb}%
+\Gamma_{dd})(1-\eta)}\right]
\end{equation}

The desired resonance (second term in the bracket) contains an $\eta$
prefactor and its width scales as $1-\eta$. When the fluctuations are
uncorrelated $(\eta=0)$ Eqs.~(\ref{Eq:am}), (\ref{Eq:av}) and (\ref{Eq:an})
can be factorized and we expect no such resonances. \ Note that these
resonances never show up in Eq. (\ref{Eq:ap}) but for $\eta=0$ cancel by
interference in Eq.~(\ref{Eq:long}). For finite $\eta$ and fast fluctuations
Eq.~(\ref{Eq:long}) can be factorized. Here, these resonances show up
naturally in Eq.~(\ref{Eq:long}), but in Eq.~(\ref{Eq:ap}) they come from the
breakdown of the factorization. For $\eta=1$ the resonance width vanishes
since there is no pure dephasing of the $bd$ transition Eq.~(\ref{Eq:bb}).
Such resonances have been observed both for collisional broadening in atomic
vapors [15] and, for phonon broadening in mixed molecular crystals [16] and
were denoted "dephasing induced" [15-17]. The present calculation shows that
more precisely they are induced by the correlations of fluctuations rather
than the fluctuations themselves, and are associated with specific
factorizations of the multipoint correlation functions.

\bigskip

\textbf{VI.\qquad DISCUSSION}

When nonlinear response functions are calculated using the density matrix, the
$n^{\prime}th$ order susceptibility has $2^{n}n!$ terms. These represent
$2^{n}$ Liouville space pathways which keep track of the complete
time-ordering of the various interactions with the bra and the ket, combined
with the $n!$ permutations of $n$ frequencies representing all possible
time-ordered interactions with the various fields. A wavefunction loop
calculation keeps track of time ordering only partially (relative time
ordering of ket and bra interactions is not maintained). This gives $n+1$
terms, times the same $n!$ permutations for a total of $(n+1)!$ terms. This
considerable reduction in the number of terms is very convenient for the
frequency-domain response, where the bookkeeping of time ordering is not
necessary anyhow.

When all field frequencies are tuned off resonance, we can neglect the
imaginary part of the Green's function. \ We can then set $\mathcal{G}%
$=$\mathcal{G}^{\dag}$\ and use forward-only propagation. Eq.~(\ref{Eq:an})
then assumes a more symmetric form%
\begin{equation}
\chi^{(3)}(-\omega_{s},\omega_{1},\omega_{2},\omega_{3})=\sum_{p_{4}%
}\left\langle V_{L}\mathcal{G(}\Omega_{3})V_{L}\mathcal{G}(\Omega_{2}%
)V_{L}\mathcal{G}(\Omega_{1})V_{L}\right\rangle \delta(\Omega_{1}+\Omega
_{2}+\Omega_{3}+\Omega_{4})
\end{equation}
p$_{4}$ denotes the summation over all 4! permutations of $\Omega_{1}%
,\Omega_{2},\Omega_{3},\Omega_{4}$ with $\omega_{1},\omega_{2},\omega
_{3},-\omega_{s}$. Diagrams (a) (b) (c) and (d) correspond to the permutations
$-\omega_{s}=\Omega_{4},\Omega_{3},\Omega_{2}$ and $\Omega_{1}$ respectively.
\ Now we have a single basic term with $(n+1)$! \ permutations, as opposed
Eq.~(\ref{Eq:an}) where we have $(n+1)$ terms each containing $n!$ permutations[18].

The loop expansion is most adequate for many-body perturbation theory and
involves a combination of forward and backward time evolution periods in
Hilbert space. The density matrix calculation, in contrast, only requires a
forward propagation, but it must be done in Liouville space. \ The time-domain
response functions may be obtained by a three-fold Fourier transform of
$\chi^{(3)}$%
\begin{equation}
S^{(3)}(t_{3},t_{2},t_{1})=\theta(t_{3})\theta(t_{2})\theta(t_{1})\int\int\int
d\omega_{1}d\omega_{2}d\omega_{3} \label{Eq:ax}%
\end{equation}

\[
\chi^{(3)}(-\omega_{s};\omega_{1},\omega_{2},\omega_{3})exp[-i\omega_{1}%
(t_{1}+t_{2}+t_{3})-i\omega_{2}(t_{2}+t_{3})-i\omega_{3}t_{3}]
\]

By substituting Eq.~(\ref{Eq:an}) in Eq.~(\ref{Eq:ax}) we can calculate the
response function by transforming the compact frequency-domain expression
obtained by a diagrammatic expansion on the Keldysh loop.

Acknowledgement

The support of the Chemical Sciences, Geosciences and Biosciences Division,
Office of Basic Energy Sciences, Office of Science, U.S. Department of Energy
is gratefully acknowledged. \ I wish to thank Christoph Marx for the careful
reading of the manuscript and useful comments.

\bigskip

[1] H. Haug and A-P. Jauho, \textit{Quantum Kinetics in Transport and Optics
of Semiconductors} (Springer-Verlag, Berlin,\ Heidelberg, 1996).

[2] L. P. Kadanoff and G. Baym, \textit{Quantum Statistical Mechanics. Green's
Function Methods in Equilibrium and Nonequilibrium Problems} (Benjamin,
Reading, MA., 1962).

[3] R. Mills, \textit{Propagators for many-particle systems; an elementary
treatment} (New York, Gordon and Breach 1969).

[4] J. Negele and H. Orland, "Quantum Many Particle Systems", (Westview Press; 1998).

[5] J. Rammer, \textit{Quantum Field Theory of Non-equilibrium States}, (Cambridge, New York, 2007).

[6] L. V. Keldysh, Sov. Phys. JETP \textbf{20}, 1018 (1965).

[7] J. Schwinger, J. Math. Phys. \textbf{2}, 407 (1961).

[8] S. Mukamel, \textit{Principles of Nonlinear Optical Spectroscopy} (Oxford
University Press, New York, 1995).

[9] N. Bloembergen, \textit{Nonlinear optics} (Benjamin, New York, 1965).

[10] S. Mukamel, Phys. Rev. E. \textbf{68}, 021111, (2003).

[11] U. Fano, Rev. Mod. Phys. \textbf{29}, 74 (1957).

[12] A. Ben-Reuven, Adv. Chem. Phys. \textbf{33}, 235 (1975).

[13]V. Chernyak, N. Wang and S. Mukamel, Physics Reports \textbf{263}, 213 (1995).

[14]D. Abramavicius and S. Mukamel, Chem. Rev., \textbf{104}, 2073 (2004).

[15] A.R. Bogdan, M.W. Downer, and N. Bloembergen, Phys. Rev. A \textbf{24}%
,623(1981);L.J. Rothberg and N. Bloembergen, Phys. Rev.A \textbf{30},820
(1984); L. Rothberg in Progress in Optics, Vol.\textbf{24}, E.Wolf, Ed.
(North-Holland, Amsterdam, 1987), p.38.

[16] J. R. Andrews and R.M. Hochstrasser, Chem. Phys. Lett. \textbf{82}, 381 (1981).

[17] R. Venkatramani and S. Mukamel, J. Phys. Chem. B \textbf{109}, 8132 (2005).

[18] J. F. Ward, Rev. Mod. Phys. \textbf{37}, 1 (1965); B. J. Orr and J. F.
Ward, Mol. Phys. \textbf{20}, 513 (1971).

\bigskip

\bigskip

\bigskip

\bigskip

\bigskip

\bigskip

\bigskip

\bigskip

\begin{figure}[p]
\includegraphics[scale=0.6,angle=-90]{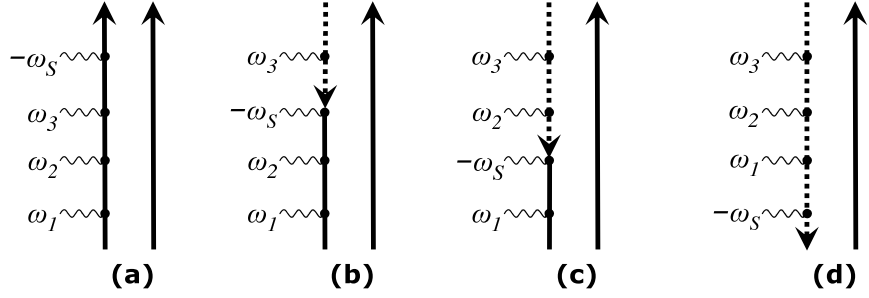}\caption{Unfolded loop diagrams
corresponding to diagrams (a), (b), (c) and (d) of Fig. 1. Eqs.(25) and (32)
can be derived directly from these diagrams using the rules given in the text.
All interactions are now from the left (ket), while the bra propagates freely.
Solid and dashed lines represent forward and backward propagation,
respectively.}%
\label{Fig:feyn2}%
\end{figure}

\begin{figure}[th]
\begin{center}
\includegraphics[scale=0.6]{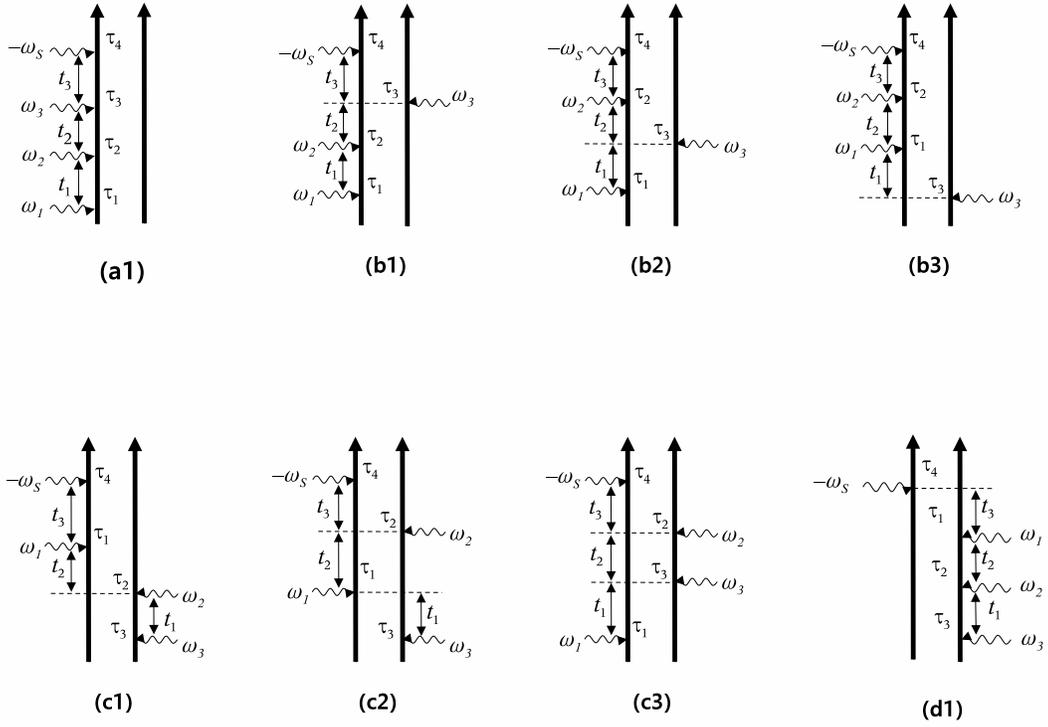}
\end{center}
\caption{The eight double-sided Feynman diagrams representing the Liouville
space pathways contributing to $\chi^{(3)}$ Eq. (28). \ Complete time ordering
of all interactions with the density matrix is maintained.$t_{1},t_{2}$ and
$t_{3}$ are the physical time intervals between successive interactions. \ (a)
and (d) of Fig.1 are time ordered and each gives only one time ordered diagram
($a_{1}$ and $d_{1})$. \ (b) and (c) of Fig.1 each split into 3 diagrams
$b_{1},b_{2}$ and $b_{3}$ and $c_{1},c_{2},c_{3}.$ \ Altogether the four loop
diagrams yield eight double-sided diagrams.}%
\label{Fig:feyn3}%
\end{figure}

\begin{figure}[p]
\includegraphics[scale=0.6,angle=-90]{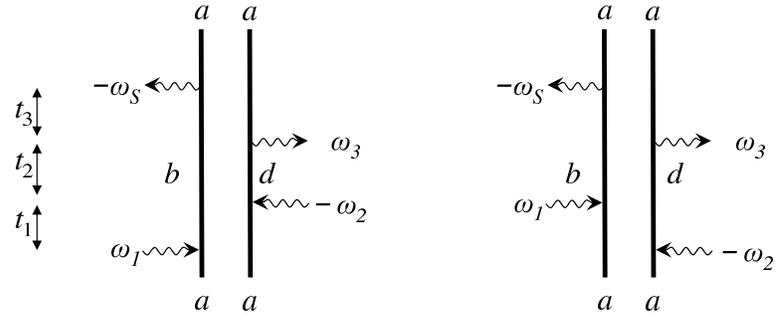}\caption{The two double-sided
diagrams which contribute to the correlation-induced resonance (Eq.44).}%
\label{Fig:feyn4}%
\end{figure}

\end{document}